**Channeling – Bent Crystals – Radiation Processes**
Frankfurt am Main, Germany
June 5–6, 2003

# Crystal Cleaning of the LHC Beam

V.M. Biryukov

Institute for High Energy Physics

142281 Protvino, Russia

**Abstract.** Channeling crystal can serve as a primary scraper for the collimation system of the Large Hadron Collider. Crystal scraper works in efficient, predictable, reliable manner with beams of very high intensity over years, and meets technical requirements imposed on the LHC collimation system. If used as primary element in the LHC collimation system, crystal would make the machine cleaner by a factor of 10-40.

## 1. Introduction

Classic two-stage collimation systems for loss localisation in accelerators typically use a small scattering target as a primary element, whereas the secondary element is a bulk absorber [1,2]. The role of the primary element is to give a substantial angular kick to incoming halo particles in order to increase the impact parameter of the particles on the absorber. An amorphous target scatters particles in all possible directions. Ideally, one would prefer a "smart target" that kicks all particles in only one direction: for instance, only in radial plane, only outward, and only into the preferred angular range corresponding to the center of absorber. Bent crystal is the first idea for such a smart target [3-7]: it traps particles and conveys them into desired direction. In physics language, we replace the scattering on single atoms of amorphous target by a coherent scattering on atomic planes of aligned crystal.

We show that bent channeling crystals can improve the collimation efficiency by a factor of 10-40 at the LHC. We start with a short historical review of the crystal channeling extraction and collimation studies and remind the recent experimental results from IHEP and RHIC. Then, we report on our Monte Carlo simulation studies for the LHC crystal collimation.

In the 1990's, an important milestone was obtained at the CERN SPS. Protons diffusing from a 120 GeV beam were extracted at an angle of 8.5 mrad with a bent silicon crystal. Efficiencies of ~10-20%, orders of magnitude higher than the values achieved previously, were measured for the first time [8-10]. The extraction studies at SPS clarified several aspects of the technique. In addition, the extraction results were found in fair agreement with Monte Carlo predictions [8,11,12]. In the late 1990's, another success came from the Tevatron extraction experiment where a crystal was channeling a 900-GeV proton beam with an efficiency of ~30% [13-16]. The simulation [17] predicted the efficiency of 35% for a realistic crystal in the Tevatron experiment. During the FNAL test, the halo created by beam-beam interaction in the periphery of the circulating beam was extracted from the beam pipe without measurable effect on the background seen by the experimental detectors.

It was predicted [11,17,18] that efficiency of crystal channeling extraction can be boosted to much higher values by multiple particle encounters with a shorter crystal installed in a circulating beam. The existence of multi-pass mechanism was confirmed in early experiments at CERN SPS and Tevatron [9,15]. To clarify this mechanism a new experiment was started at IHEP at the end of 1997, with intention to test very short crystals and achieve very high efficiencies of extraction [19,20].

## 2. The IHEP experiment on crystal collimation

In order to let the circulating particles encounter the crystal many times and suffer less scattering and nuclear interactions in the crystal, one has to minimise the crystal length down to some limit set by the physics of channeling in a strongly bent crystal [11,18]. This optimisation was studied in Monte Carlo simulations in general and for the experiments at CERN SPS and the Tevatron [11,12,17,18], taking into account the circulation of particle in the accelerator ring over many turns with multiple encounters with a bent crystal.

Such a Monte Carlo study done for the undertaken experiment at IHEP has predicted [21] that a crystal can be shorten quite a bit, down to ~1 mm along the 70-GeV beam in the extraction set-up of IHEP U-70. However, the benefits from this optimisation were expected tremendous: the crystal extraction efficiency could be as high as over 90%. Figure 1 shows both the predicted [21] dependence of the IHEP crystal extraction efficiency as a function of the crystal length, and some history of the measurements since 1997 [7,19-22].

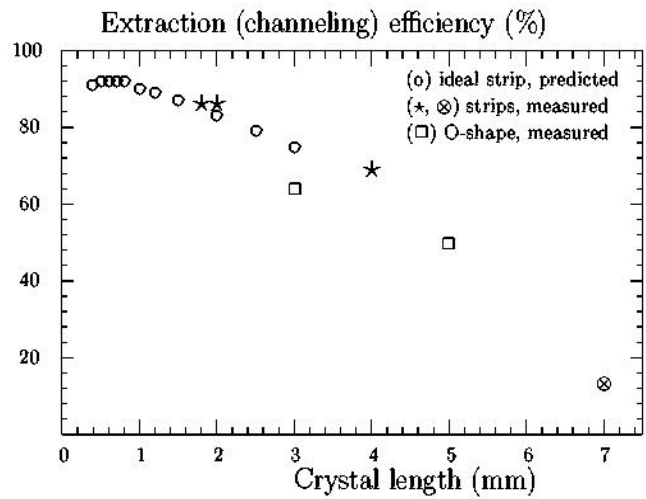

**Fig. 1** The efficiency of crystal extraction/collimation measured for 70 GeV protons as a function of the crystal length along the beam. IHEP measurements [7] and Monte Carlo prediction [21].

Producing bent crystal deflectors of required size and curvature is not an easy task, moreover as one takes into account that deflector has to be placed in a circulating beam and any extra disturbance to halo particles must be avoided. Excellent crystalline deflectors used at IHEP were produced at PNPI, IHEP, and Ferrara. In PNPI, a novel technique of "O"-shaped crystals of tiny size (Fig. 2) was invented [19] and several new approaches are being studied. In IHEP and Ferrara, "strip" -type crystals (Fig. 2) were produced which have demonstrated record 85% efficiency of channeling [7,23].

The experimentally recorded high efficiency followed nicely the prediction as seen in Fig. 1. Compared to the CERN SPS and Tevatron experiments, the efficiency is improved by a factor of 3-6 while the crystal size along the beam was reduced by a factor of 15-20 (from 30-40 mm to less than 2 mm).

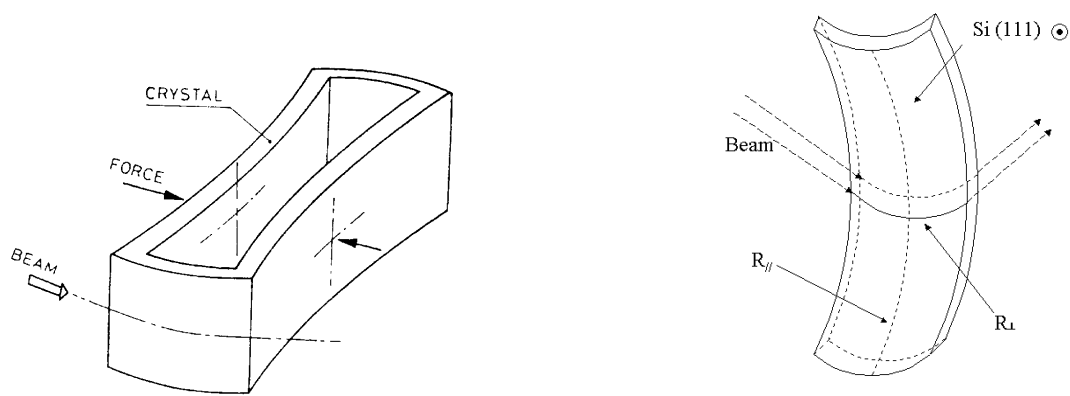

**Fig. 2** Two examples of IHEP crystals with bending devices. Left: O-shaped crystal as used in IHEP and RHIC. Right: Strip-type crystal used in IHEP.

Experimentally, the extraction efficiency was defined as the ratio of the extracted beam intensity as measured in the external beam line to all the beam loss in the circulating beam. Each measurement included the statistics from several hundred machine cycles. A remarkable feature of the IHEP extraction is that the record high efficiency of about 85% is pertained even when the *entire beam* stored in the ring is dumped onto the crystal. Figure 3 shows how the efficiency changed versus the fraction of the beam store used in the channeling experiment. Two different locations on the U-70 ring were used in the data presented in Fig. 3, straight section 106 (curve 1, o) and block 22 (curve 2, ●), and similar crystals employed. Different slopes of the experimental dependences are due to the drift of the mean incident angle when a beam is being dumped onto the crystal.

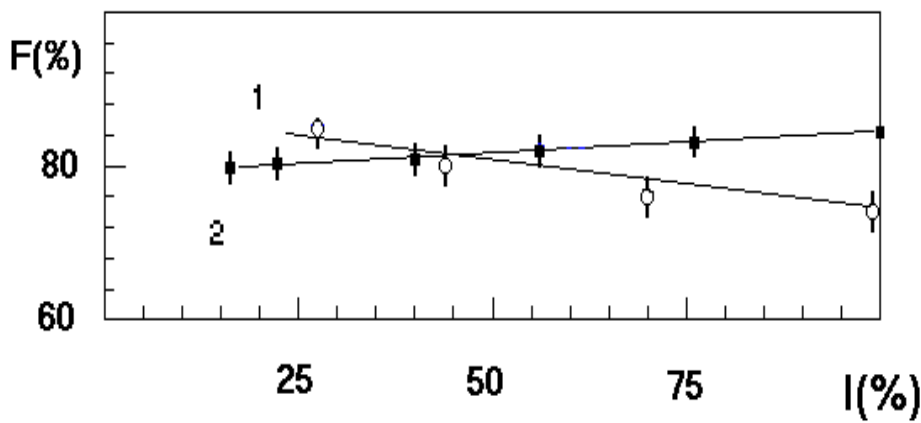

**Fig. 3** Crystal extraction efficiency as a function of the beam store fraction dumped onto the crystal. Measured on two locations (1, 2) in the U-70 ring.

IHEP has many locations on the U-70 ring where crystals are installed for extraction and collimation studies. Two of these locations are dedicated for crystal collimation. In a collimation experiment, a bent crystal is positioned upstream of a secondary collimator (stainless steel absorber 4 cm wide, 18 cm high, 250 cm long) and closer to the beam in the horizontal plane. The profilemeter records the radial distribution of the particles incident on the entry face of the secondary collimator (Fig. 4). This distribution includes the peak of channeled particles deflected into the depth of the collimator, and the nonchanneled multiply scattered particles peaked at the edge of the collimator. The efficiency figures as measured on the extraction set-up were reproduced on the collimation set-up where the intensity of the channeled beam is obtained by integration of the peak in the profile.

The collimation experiment was repeated at the injection plateau of the U-70, with 1.3 GeV protons, on the same collimation set-up with the same crystal. As one can see in Figure 4 the channeling effect is still quite profound although the energy was lowered by two orders of magnitude.

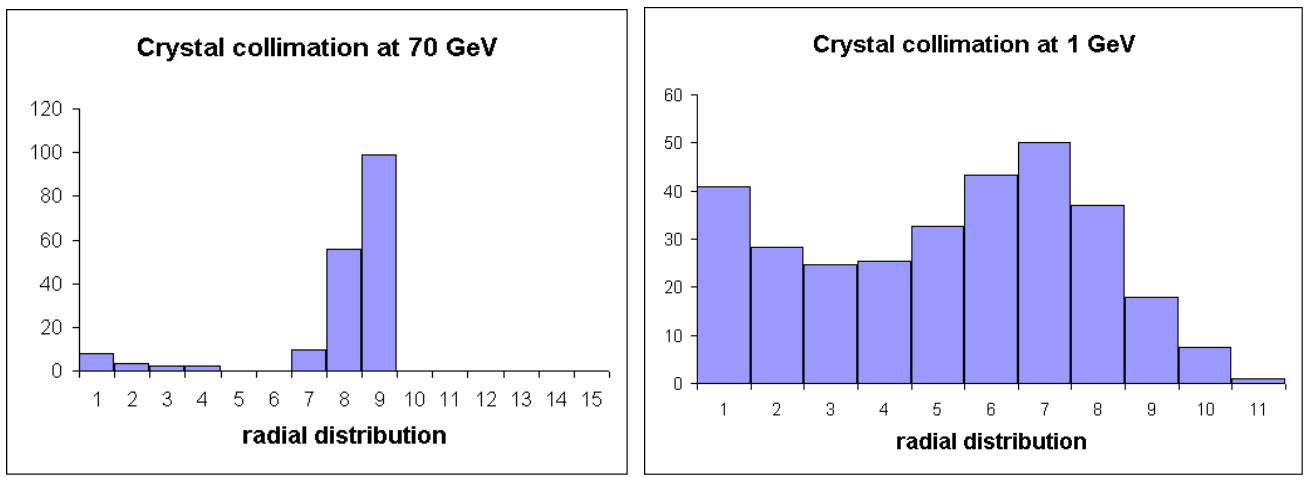

**Fig. 4** The radial beam profile observed at the entry face of the collimator with crystal working as a primary scraper, at top energy (70 GeV, left) and at injection plateau (1.3 GeV, right); crystal is the same.

## IHEP: Crystal collimation at a ramping energy

In the recent IHEP experiment the same crystal collimation set-up was tested in a broad energy range made available in the main ring of U-70 accelerator. Earlier, the experiment was performed at the top energy flattop, 70 GeV, and at the injection flattop, 1.3 GeV, of U-70 machine. This time the tests [24] were done at seven intermediate energies and, importantly, it was not possible to arrange a flattop for each energy. During the acceleration part of the machine cycle, on a certain moment corresponding to the energy of the test, the beam was dumped in a short time onto the crystal.

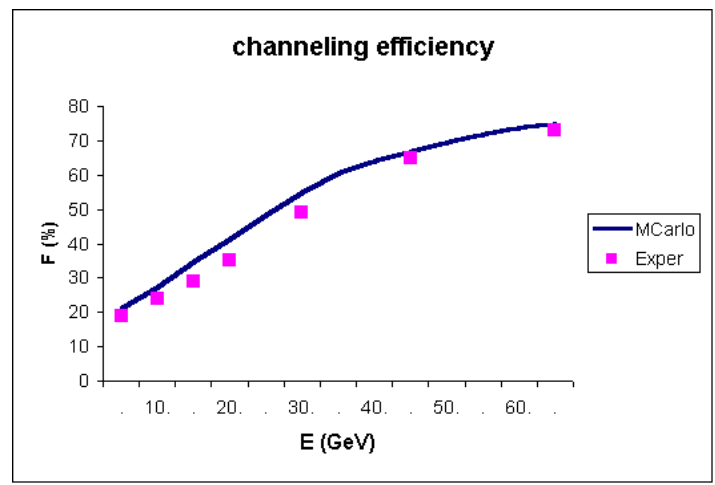

**Fig. 5** Crystal collimation efficiency (channeled particles ratio to the entire beam dump) as measured and as expected from simulation. The case of ramping energy in U-70.

These measurements are summarized in Figure 5 showing the ratio of the channeled particles to the entire beam dump (the *crystal collimation efficiency*) as measured and as predicted by Monte Carlo simulation. Figure 6 shows the examples of the radial beam profile observed at the entry face of the collimator with crystal working as primary scraper, at 12 GeV and at 45 GeV. One can see that the same crystal shows efficient work from injection through the ramping to the top energy.

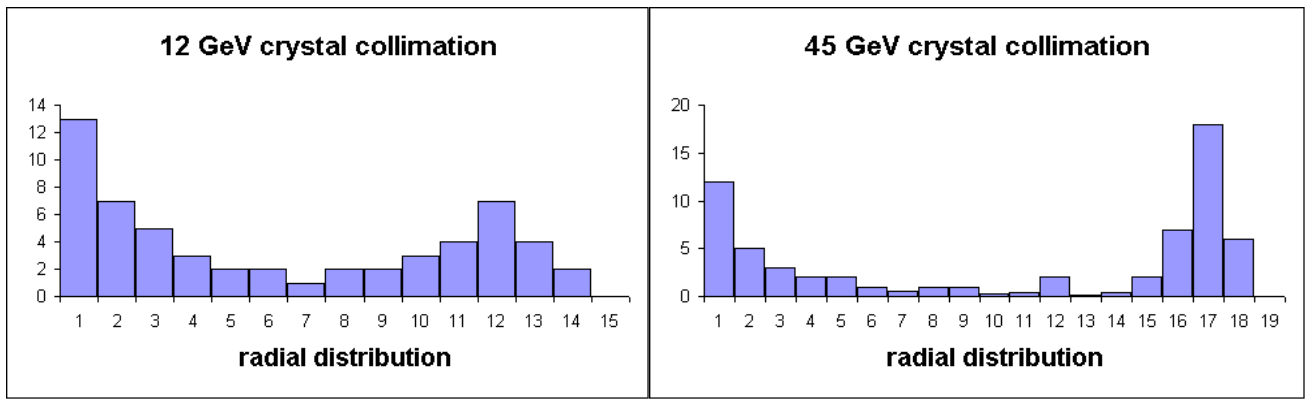

**Fig. 6** The radial beam profile observed at the entry face of the collimator with crystal working as primary scraper, at 12 GeV (left) and at 45 GeV (right).

### IHEP: Crystal collimation at high intensity

Other important issues to be addressed for a practical application of crystal-assisted extraction and collimation are thermal shock, radiation damage and crystal lifetime. In typical collimation and extraction tests at IHEP U-70, crystal channeled ~$10^{12}$ protons (up to $3 \cdot 10^{12}$ in some runs) in a spill of 0.5-1 s duration.

Let us illustrate it in the following way. Suppose, all the LHC store of $3 \cdot 10^{14}$ protons is dumped on our single crystal in a matter of 0.2 hour [1]. This makes a beam of $4 \cdot 10^{11}$ proton/s incident on the crystal face. In IHEP, this is just routine work for a crystal, practiced every day.

One of the crystals (5 mm long) located upstream of the U-70 cleaning area was exposed for several minutes to even higher radiation flux of 70 GeV protons [7]. It received ~$10^{14}$ proton hits per spill of 50 ms, with a repetition period of 9.6 s. Although it was impossible to characterise the crystal efficiency in such a short time, the channeling properties after the exposure of the crystal were tested in an external beam line. The deflected beam observed with photo-emulsion (Figure 7) was perfectly normal, without breaks, nor significant tails eventually produced by dechanneled particles. This is a good indication of the absence of thermal and radiation damages.

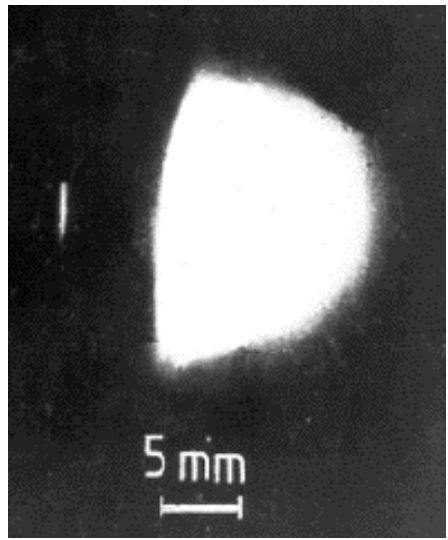

**Fig. 7** Photograph of the deflected (left) and incident (right) beams as seen downstream of the crystal. Prior to the test, the crystal was exposed in the ring to 50-ms pulses of very intense beam (~$10^{14}$ proton hits per pulse). No damage of crystal was seen in the test, after this extreme exposure.

Let us translate it to the LHC case. One bunch of the LHC is $1.1 \cdot 10^{11}$ protons. The IHEP crystal survives an instant dump of 1000 bunches of the LHC. The LHC collimation system is required to survive a hit of 20 bunches [1], so the crystal conforms to it with a great safety margin. As for the lifetime of a crystal, the CERN experiment [10] with 450 GeV protons showed that at the achieved irradiation of $5 \cdot 10^{20}$ proton/cm$^2$ the crystal lost only 30% of its deflection efficiency, which means about 100 years lifetime in the intense beam of NA48 experiment. One of the IHEP crystals served in the vacuum chamber of U-70 over 10 years, from 1989 to 1999, delivering beam to particle physicists, until a new crystal replaced it (in order to reduce the size of the channeled beam).

## 3. RHIC experiment on crystal collimation

Another experiment on crystal collimation has been in progress at the Relativistic Heavy Ion Collider [25,26]. The yellow ring of the RHIC has a bent crystal collimator of the same type (O-shaped, Figure 2) as used in earlier IHEP experiments, 5 mm along the beam. By properly aligning the crystal to the beam halo, particles entering the crystal are deflected away from the beam and intercepted downstream in a copper scraper. The purpose of a bent crystal is to improve the collimation efficiency as compared to a scraper alone.

Beam losses were recorded by the PIN diodes, hodoscope, and beam loss monitors. Signals from the RHIC experiments were also logged to monitor their background rates. Fig. 8 shows a typical angular scan from the 2003 RHIC run with gold ions. The same Figure 8 shows the predicted angular scan. The simulation is done with the measured machine optics. The two angular curves, measured and predicted, are in reasonable agreement.

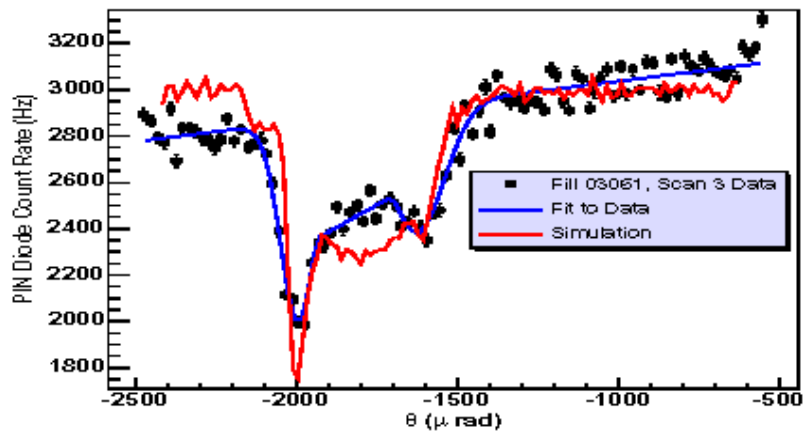

**Fig. 8** RHIC: Crystal collimator efficiency as a function of the crystal angle, for gold ions. Measured data and simulation (CATCH) [26].

The efficiency is defined as maximum depth of the large dip divided by the background rate. For the 2003 RHIC run, the theory predicted the efficiency of 32%, and averaging over the data for this run gives the measured efficiency of 26%.

The modest figure of efficiency ~30%, both in theory and experiment, is attributed to the high angular spread of the beam that hits the crystal face as set by machine optics. It is worth to compare this figure of efficiency for gold ions at RHIC to the 40% efficiency achieved with similar crystal for protons at IHEP in 1998 [19]. It is also worth to notice that the crystal extraction efficiency observed at CERN SPS with lead ions was 4 to 11% with a long (40 mm) crystal of silicon [27].

## 4. The Tevatron simulations on crystal collimation

A possibility to improve the Tevatron beam halo scraping using a bent channeling crystal instead of a thin scattering target as a primary collimator was studied at Fermilab [6]. In order to evaluate the efficiency of the collimation system, realistic simulations have been performed using the CATCH, STRUCT and MARS Monte Carlo codes.

|     | with target | with crystal amorphous layer thickness | | |
| --- | --- | --- | --- | --- |
|     |     | $10\,\mu m$ | $5\,\mu m$ | $2\,\mu m$ |
| D0  | 11.5 | 1.35 | 1.60 | 1.15 |
| CDF | 43.6 | 5.40 | 3.20 | 3.43 |
| $N$ | 270 | 82.4 | 70.6 | 50.3 |

**Table 1** The Tevatron: Halo hit rates at the D0 and CDF Roman pots and nuclear interaction rates $N$ (in $10^4$ p/s) in the primary scraper (target or crystal). Simulation [6]. Ten-fold improvement is expected from a crystal scraper.

It was shown that the scraping efficiency can be increased by one order of magnitude. As a result, the beam-related backgrounds in the CDF and D0 collider detectors can be reduced by a factor of 7 to 14. Calculated results on the system performance taking into account the thickness of near-surface amorphous layer of the crystal are presented in Table 1. Two cases have been compared:

1. The Tevatron RUN-II collimation system with all secondary collimators in design positions, but only one (D17h) horizontal primary collimator in working position. This primary collimator intercepts large amplitude protons and protons with positive momentum deviations.

2. The same collimation scheme, but silicon bent crystal is used instead of primary collimator.

## 5. Simulations of the LHC crystal collimation

We have applied the same computer model [28] as been used for the IHEP, CERN SPS, Tevatron, and RHIC experiments in order to evaluate the potential effect of crystal collimation for the LHC. In the model, a bent crystal was positioned as a primary element at a horisontal coordinate of 6σ in the halo of the LHC beam, on the location presently chosen for an amorphous primary element of the LHC collimation system design [29]. The LHC lattice functions were taken corresponding to this location.

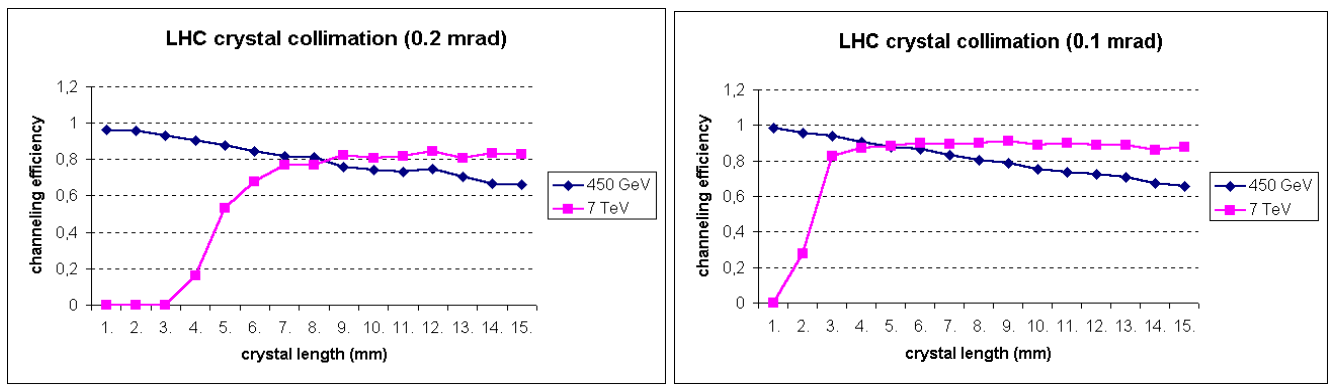

**Fig. 9** Channeling efficiency as a function of the crystal length along the LHC beam, shown for two cases: at flattop and at injection. The left figure is for the crystal bending of 0.2 mrad, the right one for 0.1 mrad.

We have varied the crystal parameters such as the size, bending, alignment angle, material, and the quality of the surface. We observed the efficiency of channeling, i.e. the number of the particles deflected at the full bending angle of the crystal, taking into account many turns in the LHC ring and multiple encounters with a crystal.

On the first encounter, the halo particles were incident at the crystal face within ≤1 micron from the edge, and the 1-micron thick near-surface layer of the crystal was assumed amorphous. I.e., on the first encounter the particles were not channeled at all, being just scattered after traversing the full crystal length. This is most conservative approach. In principle, it is possible to make a crystal face perfect to much better accuracy, like ~0.01 micron (an oxidized layer thickness) or even better. Then, many of the incident halo particles would be channeled already on the very first encounter.

Figure 9 shows the channeling efficiency as a function of the crystal length along the LHC beam for two cases: at flattop (7 TeV) and at injection (450 GeV). Different bending angles were tried. The optimal size of the silicon crystal is order of 10 mm for 0.2-mrad bending, 5 mm for 0.1 mrad, and 3 mm for 0.05 mrad. High efficiency of channeling can be obtained with the same (optimised) crystal both at 7 TeV and at 450 GeV. The efficiency becomes 90-94% in the case of crystal bending angle of 0.05-0.1 mrad.

Further, we tried different bending angles (finding every time the optimal size for the crystal) as the efficiency of channeling depends on it. Figure 10 (left) shows the channeling efficiency $F$ as a function of the crystal bending angle. Figure 10 (right) shows the same data plotted as a "background reduction factor" *1/(1-F)*. If all channeled particles were fully intercepted by the secondary collimator, then only nonchanneled particles contribute to the background in the accelerator. The factor *1/(1-F)* shows how much cleaner the accelerator could be made.

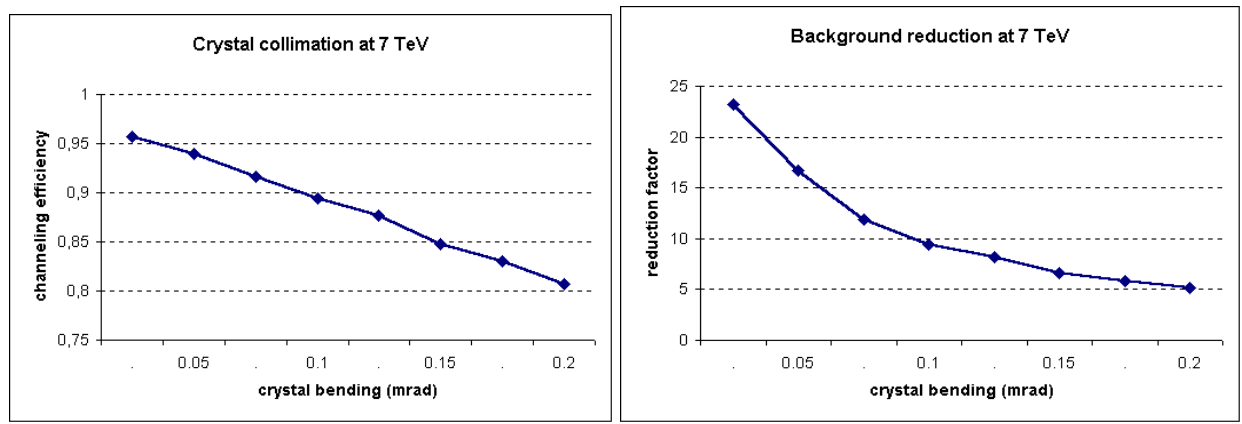

**Fig. 10** The left plot is the channeling efficiency $F$ as a function of the crystal bending angle. The right plot is the same data plotted as an LHC background reduction factor *1/(1-F)*. Silicon (110) crystal with a rough (1 micron) surface.

It should be said that all the range of crystal deflector size assumed in Fig. 10 is already realised and tested in 70 GeV beam. This technique is available and well established.

## Low-Z and high-Z primary scrapers for the LHC

The optics of traditional (amorphous) collimation at accelerators and technical considerations may require primary scrapers of different material (atomic number Z). As the technique of bent crystal channeling is developed also with other materials, e.g. germanium (Z=32) [30] and diamond (Z=6) [31], we continued our studies with other crystals.

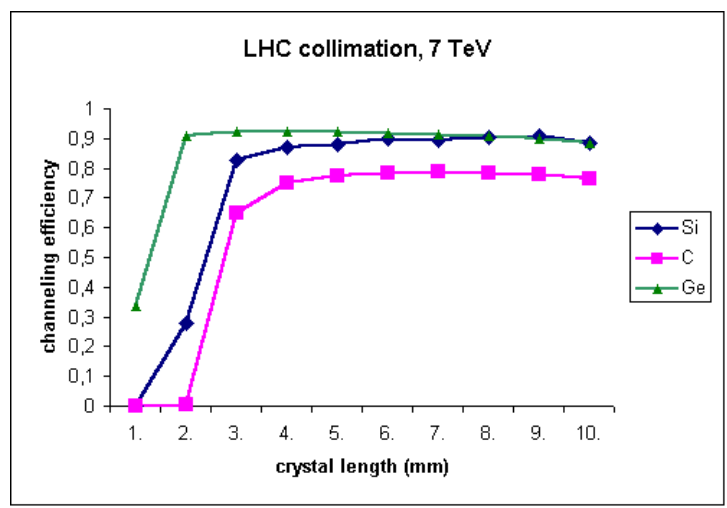

**Fig. 11** Channeling efficiency at 7 TeV as a function of the crystal length along the LHC beam, shown for different crystals: Silicon (Z=14), Diamond (Z=6), and Germanium (Z=32).

Fig. 11 gives the channeling efficiency for different crystals. Crystal plane (110) and bending 0.1 mrad were used in each case. We see that comparable efficiencies can be obtained in all these cases. All these crystals, from diamond to germanium, can serve as an LHC primary scraper. Another interesting (although futuristic) possibility might be the use of nanostructured material [32].

## 6. Conclusion

Crystal would be very efficient in the LHC environment. The expected efficiency figure, ~90%, is already experimentally demonstrated at IHEP and confirmed by simulations for the Tevatron. This will make the LHC 10 times (up to ~40 times) cleaner. Monte Carlo model successfully predicts the crystal work in the circulating beam, as demonstrated recently in crystal collimation experiments at IHEP and RHIC, and in crystal extraction experiments at up to 900 GeV (the Tevatron).

Crystal works efficiently at very high intensities (~$10^{12}$), actually much higher than the LHC requires, with a lifetime of many years. Crystal survives the abnormal dump of the LHC beam with ~100-fold safety margin (i.e. survives the instant dump of 1000 LHC bunches or ~$10^{14}$ protons) as demonstrated experimentally at 70 GeV.

The same crystal scraper works efficiently over full energy range, from injection through ramping up to top energy, as demonstrated experimentally at IHEP from 1 through 70 GeV and as seen in simulations for the LHC. Bent crystals of low-Z and high-Z material are available, e.g. diamond and germanium, and they demonstrate the efficiency similar to that of silicon. Even when a crystal is misaligned, nonchanneling, it still works as an amorphous scatterer so the collimation system returns to its traditional scheme. This makes it safe.